# Laser Spectroscopic Determination of the $^6$He Nuclear Charge Radius

L.-B. Wang,[1,2] P. Mueller,[1] K. Bailey,[1] G. W. F. Drake,[3] J. P. Greene,[1] D. Henderson,[1] R. J. Holt,[1] R. V. F. Janssens,[1] C. L. Jiang,[1] Z.-T. Lu,[1,4 *] T. P. O'Connor,[1] R. C. Pardo,[1] K. E. Rehm,[1 #] J. P. Schiffer,[1,4] X. D. Tang[1]

[1]*Physics Division, Argonne National Laboratory, Argonne, Illinois 60439*
[2]*Physics Department, University of Illinois at Urbana-Champaign, Urbana, Illinois 61801*
[3]*Department of Physics, University of Windsor, Windsor, Ontario, Canada N9B 3P4*
[4]*The Enrico Fermi Institute and Department of Physics, The University of Chicago, Chicago, Illinois 60637*

We have performed precision laser spectroscopy on individual $^6$He ($t_{1/2}$ = 0.8 s) atoms confined and cooled in a magneto-optical trap, and measured the isotope shift between $^6$He and $^4$He to be 43,194.772 ± 0.056 MHz for the $2^3S_1 - 3^3P_2$ transition. Based on this measurement and atomic theory, the nuclear charge radius of $^6$He is determined, for the first time in a method independent of nuclear models, to be 2.054 ± 0.014 fm. The result is compared with the values predicted by a number of nuclear structure calculations, and tests their ability to characterize this loosely bound, halo nucleus.

One of the most basic observables of an atomic nucleus is its size. For $^6$He ($t_{1/2}$ = 0.8 s), one of the lightest nuclear systems unstable against β-decay [1], this observable is of particular interest because two of its neutrons are loosely bound and form a 'halo' with considerably larger radial extent than the α-particle core. This halo character can be revealed by an accurate determination of the nuclear charge radius in an atomic isotope shift measurement [2] because the motion of the core with regard to the center of mass reflects both the radial extent of the neutrons and the correlations between these particles.

Recent advances in computational methods have made it possible to calculate the structure of few-nucleon (A ≤ 10) systems from the basic interactions between the constituents. *Ab initio* calculations with Monte Carlo techniques based on known two-body and empirically determined three-body potentials have demonstrated good agreement with the binding energy, spin and parity of the ground state and low-lying excited levels of all known A ≤ 10 nuclei [3]. In addition, the calculated wave functions also contain information on the spatial distribution of both protons and neutrons in these nuclei and, indeed, reproduce the charge radius in systems where this quantity is known (all radii referred to in this paper are root-mean-square radii). We have determined for the first time the charge radius of $^6$He by measuring the atomic isotope shift between $^6$He and $^4$He using laser spectroscopy. This new information on this neutron-rich nucleus is sensitive to the isospin dependence of the three-body potential, which in turn is essential to the understanding of the structure of all neutron-rich systems, including neutron stars [3].

The halo has been extensively investigated by Tanihata *et al.* [4, 5] who found that, as in $^{11}$Li, the nuclear interaction cross section of a $^6$He beam with a number of targets (C, B, etc.) was significantly larger than that of $^4$He. The effects of the halo in $^6$He were also observed in elastic scattering of protons [6]. From both measurements, an interaction radius was derived using simple models, and neutron and proton radii were obtained.

The high-precision determination of the charge radius from the present measurement of the isotope shift is made possible by recent advances in the theory of the atomic structure of helium [7]. Based on quantum mechanics with relativistic and QED corrections, a precise calculation has been performed [8], which relates the $^6$He-$^4$He isotope shifts (IS, in MHz) of both $2^3S_1 - 2^3P_2$ and $2^3S_1 - 3^3P_2$ transitions to the difference between the mean-square charge radii (in fm$^2$):

$$IS_{2S-2P} = 34473.625(13) + 1.210(\langle r_c^2 \rangle_{He4} - \langle r_c^2 \rangle_{He6}) \text{ MHz} \quad (1)$$

$$IS_{2S-3P} = 43196.202(16) + 1.008(\langle r_c^2 \rangle_{He4} - \langle r_c^2 \rangle_{He6}) \text{ MHz}. \quad (2)$$

The above isotope shifts are dominated by mass shifts, which are of the order of a few tens of GHz, with the volume shifts being of the order of 1 MHz only. The ~ 0.01 MHz errors in both mass shifts are mainly due to an uncertainty of 0.8 keV/$c^2$ in the mass of $^6$He [8]. The charge radius of $^4$He was measured to be 1.673(1) fm in previous work based on the spectroscopy of muonic $^4$He atoms [9]. The isotope shift method has also been used to precisely determine the charge radius of $^3$He (= 1.9506 (14) fm) [10], a result consistent with the less precise values obtained from electron scattering on a $^3$He target.

The challenges presented by the high precision required in the laser spectroscopy measurement and the high sensitivity needed to probe the small number of $^6$He atoms available have led us to the approach of performing laser spectroscopy on individual $^6$He atoms confined and cooled in a magneto-optical trap (MOT). In this work, $^6$He nuclei were produced in a hot (750°C) graphite target via the $^{12}$C($^7$Li, $^6$He)$^{13}$N reaction with a 100 pnA, 60 MeV beam of $^7$Li from the ATLAS accelerator at Argonne National Laboratory. Neutral $^6$He atoms diffused out of the target and were transferred in vacuum to the nearby atomic beam assembly in approximately 1 s. By detecting the characteristic β-decay, we established that $^6$He atoms were transferred to the atomic beam assembly at the rate of ~ 1 × 10$^6$ s$^{-1}$. Details on the production and transfer of $^6$He atoms are given in [11]. Our design of the atomic beam assembly is based on a type of MOT system widely used to trap various metastable noble gas atoms [12]. Trapping helium atoms in the $2^3S_1$ metastable level was accomplished by exciting the $2^3S_1 - 2^3P_2$ transition using laser light with a wavelength of 1083 nm generated by a system consisting of a diode laser and a fiber amplifier. In the experiment, $^6$He atoms were mixed with a krypton carrier gas and sent through a ceramic tube of 1 cm diameter within which a RF-induced discharge was used to excite a fraction (~ 10$^{-5}$) of the $^6$He atoms to the $2^3S_1$ level. The metastable atoms were transversely cooled, then decelerated with the Zeeman slowing technique and captured in a MOT. $^6$He atoms remained trapped for an average of only 0.4 s due to β-decay and collisional losses. This trap system captured $^6$He atoms at a typical rate of 100 hr$^{-1}$, with a total capture efficiency of 2 × 10$^{-8}$. For the detection and spectroscopy of a single trapped atom, we chose to excite the $2^3S_1 - 3^3P_2$ transition at a wavelength of 389 nm mainly because photons of this wavelength can be detected using a photomultiplier tube with adequate efficiency (~ 25%). In the trap, a single atom emitted resonant photons at a rate of $10^5$ s$^{-1}$, of which 0.5% were collected and counted. Fluorescence from a single atom induced a photon count rate of $7 \times 10^2$ s$^{-1}$ while the background due to photons scattered off the walls was at $2 \times 10^2$ s$^{-1}$. A single trapped $^6$He atom was identified in 0.1 s with a signal-to-noise ratio of 15 (Fig. 1).

The 389 nm light was generated through frequency doubling of the amplified output of an external-cavity diode laser (DL1) at 778 nm. The frequency of DL1, after being shifted by a tunable acousto-optical modulator, was locked to a Fabry-Perot interferometer (FPI). The frequency mode of the FPI was locked to a second diode laser at 778 nm (DL2), whose frequency in turn was locked to a saturation absorption peak of an $I_2$ molecular transition. This particular transition has a linewidth of 5 MHz, and was chosen because its frequency is within the scan range covering both $^4$He and $^6$He spectroscopy. The beat frequency between DL1 and DL2 was continuously monitored using a microwave frequency counter with an accuracy of better than 0.3 parts per million. The frequency stability of this $I_2$-based spectrometer is of critical importance for the accuracy of the final measurement, and was tested extensively by repeated spectroscopy measurements on $^4$He. The statistical error in the frequency determination based on this spectrometer is 0.1 MHz with an integration time of one minute.

In the experiment, most of the time was spent in the *Capture* phase, waiting for a $^6$He atom to be captured. During this phase, the trapping laser beams (at 1083 nm) were constantly on, with their intensity (10 mW/cm$^2$ for each beam) and frequency detuning (- 20 MHz) tuned to maximize the capture probability; the probing laser beams (at 389 nm) were also constantly on, with their frequency tuned to the modified resonance (by the light shift due to trapping light) in order to maximize the fluorescence signal. Within 0.1 s of a $^6$He atom entering the trap, it was identified and the system was switched to the *Spectroscopy* phase. During the latter phase, the trapping laser beams had a lower intensity (0.8 mW/cm$^2$) and a smaller frequency detuning (-3 MHz) in order to provide a tighter confinement and more cooling to the trapped atom. In addition, the trapping laser and the probing laser beams were chopped alternately at 100 kHz. For each chopping period of 10 μs, the trapping laser beams were on for 8 μs to re-capture and cool the atom, and the probing laser beams were on for 2 μs during which the fluorescence was collected. Meanwhile, the frequency of the probing laser was scanned over a range of 18 MHz at a repetition rate of approximately 85 kHz. The fast scan and switches, performed with a set of acousto-optical modulators, were required to minimize systematic effects due to the heating/cooling of the atom by the probing light. Furthermore, in order to minimize the dependence of the spectrum on the magnetic field and on the intensity of the probing laser beams, the latter beams were linearly polarized, and the intensities of the two counter-propagating probing beams were carefully balanced.

All these controls and adjustments were tested extensively off-line by performing laser spectroscopy on $^4$He atoms. Figure 2(a) shows a typical spectrum on the $2^3S_1 - 3^3P_2$ transition accumulated over five minutes with a trap filled with a few $^4$He atoms. Measurements on this transition, as well as on the $2^3S_1 - 3^3P_0$ and $2^3S_1 - 3^3P_1$ transitions, were performed repeatedly while changing the intensity of the probing beams by as much as a factor of 60 and changing the magnetic field gradient of the MOT by a factor of two. The standard deviation of 30 measurements performed under different probing laser intensities and magnetic field gradients was 0.040 MHz, and represents the systematic error in the isotope shift measurement due to trap effects (Table 1), which include any residual Zeeman shifts and the effects of heating/cooling of the atom. The fine structure splittings of the $3^3P$ level of $^4$He measured in this manner are in agreement with the calculated values within the systematic error [13]. Moreover, the isotope shift between the $2^3S_1 - 3^3P_2$ transition of $^4$He and the $(2^3S_1, F = 3/2) - (3^3P_2\ F = 5/2)$ transition of $^3$He was measured, and the result agrees with that of Ref. [14] within the ~ 0.150 MHz error of the previous measurement. Figure 2(b) presents a spectrum on the $2^3S_1 - 3^3P_2$ transition accumulated over one hour with a total of 150 trapped $^6$He atoms. The center frequencies of the $^6$He and $^4$He spectra were obtained through fits with a Gaussian function, and the difference in the two values, after correcting for the recoil effect (Table 1), was taken as the isotope shift between $^6$He and $^4$He. A total of 18 such measurements with comparable precision, performed during two separate runs one month apart, achieved statistically consistent results (reduced chi-square = 0.85), corresponding to a statistical error of 0.033 MHz (Table 1). Other significant systematic errors included a contribution due to background variations over the scanned spectrum (0.020 MHz) and an uncertainty in the microwave frequency measurement (0.009 MHz). The light-shift effect due to the incomplete extinction of the trapping light was negligible (< 0.001 MHz). Based on the weighted average of our 18 independent measurements, the isotope shift between $^6$He and $^4$He on the $2^3S_1 - 3^3P_2$ transition was determined to be 43,194.772 ± 0.056 MHz. According to Eq. (2), this translates into a difference between the mean-square charge radii $<r_c^2>_{He6} - <r_c^2>_{He4}$ of 1.418 ± 0.058 fm$^2$. With the previously determined charge radius of $^4$He (1.673(1) fm) [9], the charge radius of $^6$He from the present measurements is then 2.054 ± 0.014 fm.

In nuclear structure theories, the spatial distributions of protons and neutrons are calculated while treating both as point particles. The point-proton radius ($<r_p^2>^{1/2}$) is related to the charge radius ($<r_c^2>^{1/2}$) by the relation:

$$<r_p^2> = <r_c^2> - <R_p^2> - <R_n^2> (N/Z), \qquad (3)$$

where $<R_p^2>^{1/2}$ (= 0.895(18) fm) [15] is the charge radius of the proton, $<R_n^2>$ (= − 0.120(5) fm$^2$) [16] is the mean-square-charge radius of the neutron, and N and Z are the neutron and proton numbers. Using Eq. (3), we derive the point-proton radius of $^6$He to be 1.912 ± 0.018 fm.

Figure 3 compares the experimental and theoretical values of the point-proton radius of $^6$He. The two earlier experimental values were extracted from nuclear collision measurements. The interpretation of such data requires both a description of the interaction and a model for the nucleon distribution in $^6$He. The value obtained in this work represents the first model-independent determination. It has achieved a much improved accuracy, and is in disagreement with that previously derived from the interaction cross section [5], presumably reflecting the inadequacies of the model assumptions.

The point-proton radius of $^6$He has been calculated using a variety of cluster models. Some describe $^6$He with the ($\alpha + n + n$) channel alone [17-19], while others include the additional ($t + t$) channel [20, 21]. Among these, both Funada *et al*. [17] and Esbensen *et al*. [19] predicted the radius to be 1.88 fm, within 2% of our experimental value. The remaining cluster model calculations under-predict the radius. The prediction by the *ab initio* calculations based on the no-core shell model [22] differs from the experimental value by 0.15 fm, or eight times the experimental uncertainty. In contrast, the value predicted by the *ab initio* quantum Monte-Carlo calculations based on the AV18 two-body potential and IL2 three-body potential agrees with our experimental value, while that obtained using another three-body potential (UIX) over-predicts the radius [23]. A new calculation using the quantum Monte-Carlo method is underway [24].

We are grateful to R. Wiringa and S. Pieper for stimulating discussions that helped initiate this project. We thank P. Collon, X. Du, A. M. Heinz, C. Law, I. D. Moore, M. Paul and the late T. Pennington for their contributions in the early stages of the project. We thank J. Michael for lending us a critical laser. This work is supported by the U.S. Department of Energy, Office of Nuclear Physics, under contract W-31-109-ENG-38. G. Drake acknowledges support by NSERC and by SHARCnet.

**Fig. 1.** The fluorescence signal of a single trapped metastable $^6$He atom. The count rate of the 389 nm fluorescence photons emitted from a single trapped atom is 35 counts in 50 ms, or $7 \times 10^2$ s$^{-1}$; the rate of background photons scattered off the walls and windows of the trap chamber is $2 \times 10^2$ s$^{-1}$.

**Fig. 2.** Laser spectroscopy of helium atoms in the MOT. Fluorescence is recorded while the probing laser frequency is scanned over the resonance of the $2^3S_1 - 3^3P_2$ transition. (a) Spectrum of $^4$He accumulated with a total approximately 1,000 atoms in 5 minutes. The best fit with a Gaussian function gives a statistical error of 0.029 MHz in the center frequency, a FWHM of 6.8 ± 0.1 MHz, and a reduced chi-square of 0.77. (b) Spectrum of $^6$He accumulated with approximately 100 atoms in one hour. The best fit with a Gaussian function gives a statistical error of 0.111 MHz in the center frequency, a FWHM of 6.2 ± 0.4 MHz, and a reduced chi-square of 1.1.

**Fig. 3.** Experimental (the top three data points) and theoretical (all remaining points) values of the point-proton radius of $^6$He. Ref. [5] was determined in nuclear reaction measurements, while [6] was extracted from elastic scattering on protons. Ref. [17-21] are results of calculations with cluster models. Ref. [22] was calculated using the no-core shell model, and [23] refers to the quantum Monte-Carlo technique described in the text with the three displayed values corresponding to different potentials between nucleons.

**Table 1.** Errors and corrections in the isotope shift of the $2^3S_1 - 3^3P_2$ transition between $^6$He and $^4$He. The isotope shift determined in this work is 43,194.772 ± 0.056 MHz.

| Source | Correction (MHz) | Error (MHz) |
|---|---|---|
| Statistical | | 0.033 |
| Trap effects | | 0.040 |
| Uneven background | | 0.020 |
| Frequency counter | | 0.009 |
| Recoil effect | + 0.110 | < 0.001 |
| Total | + 0.110 | 0.056 |

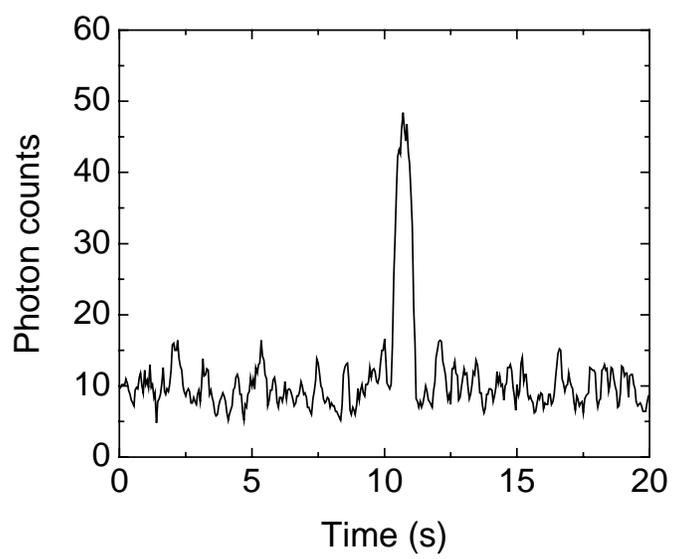 Fig. 1

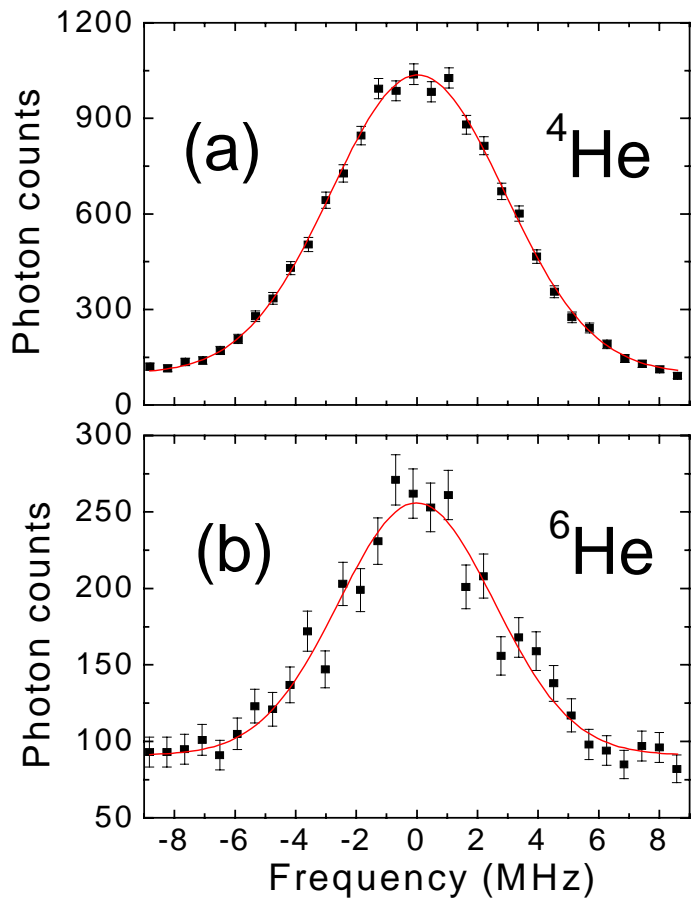

Fig. 2

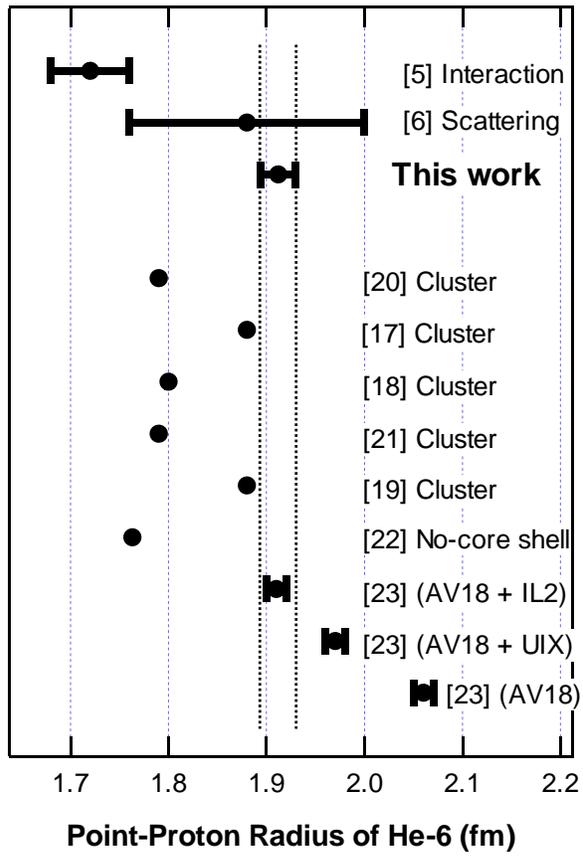

Fig. 3